\documentclass[final,5p,times,twocolumn]{elsarticle}

\usepackage{amsmath}
\usepackage{amssymb}
\usepackage{graphicx}
\usepackage{epsfig}

\journal{Physics Letters B}

\begin{document}

\begin{frontmatter}

\title{The self-consistent general relativistic solution for a system of degenerate neutrons, protons and electrons in $\beta$-equilibrium}

\author[roma,pescara]{M. Rotondo}
\ead{michael.rotondo@icra.it}

\author[roma,pescara]{Jorge A. Rueda}
\ead{jorge.rueda@icra.it}

\author[roma,pescara,nizza]{R. Ruffini}
\ead{ruffini@icra.it}

\author[roma,pescara]{S.-S. Xue}
\ead{xue@icra.it}
   
\address[roma]{Dipartimento di Fisica and ICRA, Sapienza Universita' di Roma\\
P.le Aldo Moro 5, I-00185 Rome, Italy}
\address[pescara]{ICRANet, P.zza della Repubblica 10, I-65122 Pescara, Italy}
\address[nizza]{ICRANet, University of Nice-Sophia Antipolis, 28 Av. de Valrose, 06103 Nice Cedex 2, France}

\begin{abstract}
We present the self-consistent treatment of the simplest, nontrivial, self-gravitating system of degenerate neutrons, protons and electrons in $\beta$-equilibrium within relativistic quantum statistics and the Einstein-Maxwell equations. The impossibility of imposing the condition of local charge neutrality on such systems is proved, consequently overcoming the traditional Tolman-Oppenheimer-Volkoff treatment. We emphasize the crucial role of imposing the constancy of the generalized Fermi energies. A new approach based on the coupled system of the general relativistic Thomas-Fermi-Einstein-Maxwell equations is presented and solved. We obtain an explicit solution fulfilling global and not local charge neutrality by solving a sophisticated eigenvalue problem of the general relativistic Thomas-Fermi equation. The value of the Coulomb potential at the center of the configuration is $eV(0)\simeq m_\pi c^2$ and the system is intrinsically stable against Coulomb repulsion in the proton component. This approach is necessary, but not sufficient, when strong interactions are introduced.
\end{abstract}


\end{frontmatter}

\section{Introduction}

The insurgence of critical electric fields in the process of gravitational collapse leading to vacuum polarization process \cite{physrep} has convinced us of the necessity of critically reexamining the gravitational and electrodynamical properties in neutron stars. In this light we have recently generalized the Feynman, Metropolis and Teller treatment of compressed atoms to the relativistic regimes \cite{PRC2011}. We have so enforced, self-consistently in a relativistic Thomas-Fermi equation, the condition of $\beta$-equilibrium extending the works of V.~S.~Popov \cite{popov1}, Ya.~B.~Zeldovich and V.~S.~Popov \cite{popov2}, A.~B.~Migdal et al.~\cite{migdal76,migdal77}, J.~Ferreirinho et al.~\cite{ferreirinho80} and R.~Ruffini and L.~Stella \cite{ruffini81} for heavy nuclei. Thanks to the existence of scaling laws (see \cite{PRC2011} and \cite{dresden}) this treatment has been extrapolated to compressed nuclear matter cores of stellar dimensions with mass numbers $A\simeq (m_{\rm Planck}/m_n)^3 \sim 10^{57}$ or $M_{core}\sim M_{\odot}$. Such configurations fulfill global but not local charge neutrality. They have electric fields on the core surface, increasing for decreasing values of the electron Fermi energy $E^F_e$ reaching values much larger than the critical value $E_c = m_e^2c^3/(e\hbar)$, for $E_e^F=0$. The assumption of constant distribution of protons at nuclear densities simulates, in such a treatment, the confinement due to the strong interactions in the case of nuclei and heavy nuclei and due to both the gravitational field and the strong interactions in the case of nuclear matter cores of stellar sizes.

In this article we introduce explicitly the effects of gravitation by considering a general relativistic system of degenerate fermions composed of neutrons, protons and electrons in $\beta$-equilibrium: this is the simplest nontrivial system in which new electrodynamical and general relativistic properties of the equilibrium configuration can be clearly and rigorously illustrated. We first prove that the condition of local charge neutrality can never be implemented since it violates necessary conditions of equilibrium at the microphysical scale. We then prove the existence of a solution with global, but not local, charge neutrality by taking into account essential gravito-electrodynamical effects. First we recall the constancy of the general relativistic Fermi energy of each specie pioneered by O.~Klein \cite{klein}. We subsequently introduce the general relativistic Thomas-Fermi equations for the three fermion species fulfilling relativistic quantum statistics, governed by the Einstein-Maxwell equations. The solution of this system of equations presents a formidable mathematical challenge in theoretical physics. The traditional difficulties encountered in proving the existence and unicity of the solution of the Thomas-Fermi equation \cite{tfsol1,tfsol2,tfsol3,tfsol4,tfsol5,tfsol6,tfsol7} are here enhanced by the necessity of solving the general relativistic Thomas-Fermi equation coupled with the Einstein-Maxwell system of equations. We present the general solution for the equilibrium configuration, from the center of the star all the way to the border, giving the details of the gravitational field, of the electrodynamical field as well as of the conserved quantities. 

We illustrate such a solution by selecting a central density $\rho(0) = 3.94 \rho_{\rm nuc}$, where $\rho_{\rm nuc} \simeq 2.7\times 10^{14}$ g cm$^{-3}$ is the nuclear density. We point out the existence near the boundary of the core in the equilibrium configuration of three different radii, in decreasing order: $R_e$ corresponding to the vanishing of the Fermi momentum of the electron component; $P_e^F=0$, $R_p$ corresponding to the vanishing of the Fermi momentum of the proton component; $P_p^F=0$ and $R_n$ corresponding to the radius at which the Fermi momentum of neutrons vanishes: $P_n^F=0$. We then give explicit expressions for the proton versus electron density ratio and the  proton versus neutron density ratio for any value of the radial coordinate as well as for the electric potential at the center of the configuration. A novel situation occurs: the description of the pressure and density is not anylonger a local one. Their determination needs prior knowledge of the global electrodynamical and gravitational potentials on the entire system as well as of the radii $R_n$, $R_p$ and $R_e$. This is a necessary outcome of the self-consistent solution of the eigenfunction within general relativistic Thomas-Fermi equation in the Einstein-Maxwell background. As expected from the considerations in \cite{PRC2011}, the electric potential at the center of the configuration fulfills $eV(0) \simeq m_\pi c^2$ and the gravitational potential $1-{\rm e}^{\nu(0)/2} \simeq m_\pi/m_p$. The implementation of the constancy of the general relativistic Fermi energy of each particle species and the consequent system of equations illustrated here is the simplest possible example admitting a rigorous nontrivial solution. It will necessarily apply in the case of additional particle species and of the inclusion of nuclear interactions: in this cases however it is not sufficient and the contribution of nuclear fields must be taken into due account.

\section{The impossibility of a solution with local charge neutrality}

We consider the equilibrium configurations of a degenerate gas of neutrons, protons and electrons with total matter energy density and total matter pressure
\begin{eqnarray}
{\cal E} &=& \sum_{i=n,p,e} \frac{2}{(2 \pi \hbar)^3} \int_0^{P^F_i} \epsilon_i(p)\,4 \pi p^2 dp\, ,\label{eq:eos1}\\
P &=&\sum_{i=n,p,e} \frac{1}{3} \frac{2}{(2 \pi \hbar)^3} \int_0^{P^F_i} \frac{p^2}{\epsilon_i(p)}\,4 \pi p^2 dp \, ,\label{eq:eos2}
\end{eqnarray}
where $\epsilon_i(p) = \sqrt{c^2 p^2+m^2_i c^4}$ is the relativistic single particle energy. In addition, we require the condition of $\beta$-equilibrium between neutrons, protons and electrons
\begin{equation}\label{eq:betaeq}
\mu_n = \mu_p + \mu_e\, ,
\end{equation}
where $P^F_i$ denotes the Fermi momentum and $\mu_i = \partial {\cal E}/\partial n_i = \sqrt{c^2 (P^F_i)^2+m^2_i c^4}$ is the free-chemical potential of particle-species with number density $n_i = (P^F_i)^3/(3 \pi^2 \hbar^3)$.
We now introduce the extension to general relativity of the Thomas-Fermi equilibrium condition on the generalized Fermi energy $E^F_e$ of the electron component
\begin{equation}\label{eq:electroneq}
E^F_e = {\rm e}^{\nu/2} \mu_e - m_e c^2 - e V = {\rm constant}\, ,
\end{equation}
where $e$ is the fundamental charge, $V$ is the Coulomb potential of the configuration and we have introduced the metric 
\begin{equation}\label{eq:metric}
ds^2 = {\rm e}^{\nu(r)} c^2 dt^2 - {\rm e}^{\lambda(r)} dr^2 - r^2 d\theta^2 - r^2 \sin^2 \theta d\varphi^2\, ,
\end{equation}
for a spherically symmetric non-rotating neutron star. The metric function $\lambda$ is related to the mass $M(r)$ and the electric field $E(r) = -{\rm e}^{-(\nu+\lambda)/2} V'$ (a prime stands for radial derivative) through 
\begin{equation}\label{eq:lambda}
{\rm e}^{-\lambda} = 1 - \frac{2 G M(r)}{c^2 r} + \frac{G}{c^4} r^2 E^2(r)\, .
\end{equation}
Thus the equations for the neutron star equilibrium configuration consist of the following Einstein-Maxwell equations and general relativistic Thomas-Fermi equation
\begin{eqnarray}
&&M' = 4 \pi r^2 \frac{{\cal E}}{c^2} - \frac{4 \pi r^3}{c^2} {\rm e}^{-\nu/2} \hat{V}' (n_p-n_e),\label{eq:Gab1}\\
&&\nu' = \frac{2 G}{c^2} \frac{4 \pi r^3 P/c^2 + M - r^3 E^2/c^2}{r^2 \left(1-\frac{2 G M}{c^2 r} + \frac{G r^2}{c^4} E^2 \right)}
,\label{eq:Gab2}\\
&&P'+\frac{\nu'}{2} ({\cal E} + P) = - (P^{\rm em})' - \frac{4 P^{\rm em}}{r}\, ,\label{eq:TOV}\\
&&\hat{V}'' + \frac{2}{r}\hat{V}' \left[ 1 - \frac{r (\nu'+\lambda')}{4}\right] = - 4 \pi \alpha \hbar c \, {\rm e}^{\nu/2} {\rm e}^{\lambda} \Bigg\{ n_p \nonumber \\
&&- \frac{{\rm e}^{-3 \nu/2}}{3 \pi^2}[\hat{V}^2 + 2 m_e c^2 \hat{V} - m^2_e c^4 ({\rm e}^{\nu}-1)]^{3/2}\Bigg\}\, ,\label{eq:GRTF}
\end{eqnarray}
where $\alpha$ denotes the fine structure constant, $\hat{V} = E^F_e + eV$, $P^{\rm em}= -E^2/(8\pi)$ and we have used Eq.~(\ref{eq:electroneq}) to obtain Eq.~(\ref{eq:GRTF}).

It can be demonstrated that the assumption of the equilibrium condition (\ref{eq:electroneq}) together with the $\beta$-equilibrium condition (\ref{eq:betaeq}) and the hydrostatic equilibrium (\ref{eq:TOV}) is enough to guarantee the constancy of the generalized Fermi energy
\begin{equation}\label{eq:olsoneq}
E^F_i = {\rm e}^{\nu/2} \mu_i - m_i c^2 + q_i V \, ,\qquad i = n,p,e\, ,
\end{equation}
for all particle species separately. Here $q_i$ denotes the particle unit charge of the $i$-species. Indeed, as shown by Olson and Bailyn \cite{olson75,olson78}, when the fermion nature of the constituents and their degeneracy is taken into account, in the configuration of minimum energy the generalized Fermi energies $E^F_i$ defined by (\ref{eq:olsoneq}) must be constant over the entire configuration. These minimum energy conditions generalize the equilibrium conditions of Klein \cite{klein} and of Kodama and Yamada \cite{kodama72} to the case of degenerate multicomponent fluids with particle species with non-zero unit charge.

\begin{figure}[h!]
\centering
\includegraphics[width=\columnwidth,clip]{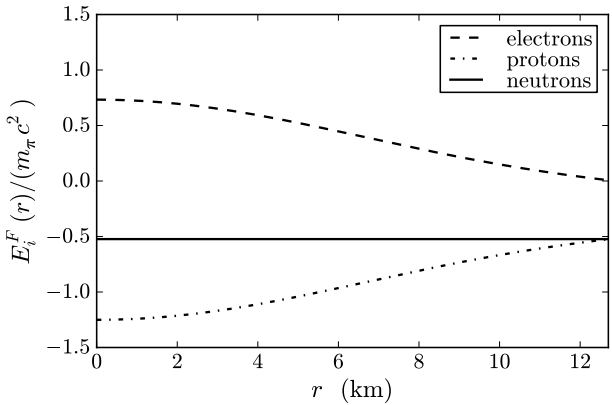}
\caption{Fermi energies for neutrons, protons and electrons in units of the pion rest-energy for a locally neutral configuration with central density $\rho(0) = 3.94 \rho_{\rm nuc}$, where $\rho_{\rm nuc} = 2.7\times 10^{14}$ g cm$^{-3}$ denotes the nuclear density.}\label{fig:1}
\end{figure}

If one were to assume, as often done in literature, the local charge neutrality condition $n_e(r) = n_p(r)$ instead of assuming the equilibrium condition (\ref{eq:electroneq}), this would lead to $V=0$ identically (since there will be no electric fields generated by the neutral matter distribution) implying via Eqs.~(\ref{eq:betaeq}) and (\ref{eq:TOV})
\begin{eqnarray}
E^F_e + E^F_p &=& {\rm e}^{\nu/2}(\mu_e + \mu_p)-(m_e+m_p) c^2 = E^F_n\nonumber \\
&+&  (m_n-m_e-m_p) c^2 = {\rm constant}\, .\label{eq:ln2}
\end{eqnarray}
Thus the neutron Fermi energy would be constant throughout the configuration as well as the sum of the proton and electron Fermi energies but not the individual Fermi energies of each component. In Fig.~\ref{fig:1} we show the results of the Einstein equations for a selected value of the central density of a system of degenerate neutrons, protons, and electrons in $\beta$-equilibrium under the constraint of local charge neutrality. In particular, we have plotted the Fermi energy of the particle species in units of the pion rest-energy. It can be seen that indeed the Fermi energies of the protons and electrons are not constant throughout the configuration which would lead to microscopic instability. This proves the impossibility of having a self-consistent configuration fulfilling the condition of local charge neutrality for our system. This result is complementary to the conclusion of Eq.~(4.6) of \cite{olson75} who found that, at zero temperature, only a dust solution with zero particle kinetic energy can satisfy the condition of local charge neutrality and such a configuration is clearly unacceptable for an equilibrium state of a self-gravitating system.

\section{The solution with global charge neutrality}

We turn now to describe the equilibrium configurations fulfilling only global charge neutrality. We solve self-consistently Eqs.~(\ref{eq:Gab1}) and (\ref{eq:Gab2}) for the metric, Eq.~(\ref{eq:TOV}) for the hydrostatic equilibrium of the three degenerate fermions and, in addition, we impose Eq.~(\ref{eq:betaeq}) for the $\beta$-equilibrium. The crucial equation relating the proton and the electron distributions is then given by the general relativistic Thomas-Fermi equation (\ref{eq:GRTF}).
The boundary conditions are: for Eq.~(\ref{eq:Gab1}) the regularity at the origin: $M(0)=0$, for Eq.~(\ref{eq:TOV}) a given value of the central density, and for Eq.~(\ref{eq:GRTF}) the regularity at the origin $n_e(0)=n_p(0)$, and a second condition at infinity which results in an eigenvalue problem determined by imposing the global charge neutrality conditions
\begin{equation}\label{eq:bound1}
\hat V (R_e) = E^F_e\, ,\qquad \hat V'(R_e) = 0\, ,
\end{equation}
at the radius $R_e$ of the electron distribution defined by 
\begin{equation}\label{eq:bound2}
P^F_e (R_e) = 0\, ,
\end{equation}
from which follows
\begin{eqnarray}\label{eq:bound3}
E^F_e &=& m_e c^2 {\rm e}^{\nu(R_e)/2} - m_e c^2 \nonumber \\ 
&=& m_e c^2 \sqrt{1-\frac{2 G M(R_e)}{c^2 R_e}} - m_e c^2\, .
\end{eqnarray}
Then the eigenvalue problem consists in determining the gravitational potential and the Coulomb potential at the center of the configuration that satisfy the conditions (\ref{eq:bound1})--(\ref{eq:bound3}) at the boundary.

\section{Numerical integration of the equilibrium equations}

\begin{figure}
\centering
\includegraphics[width=\columnwidth,clip]{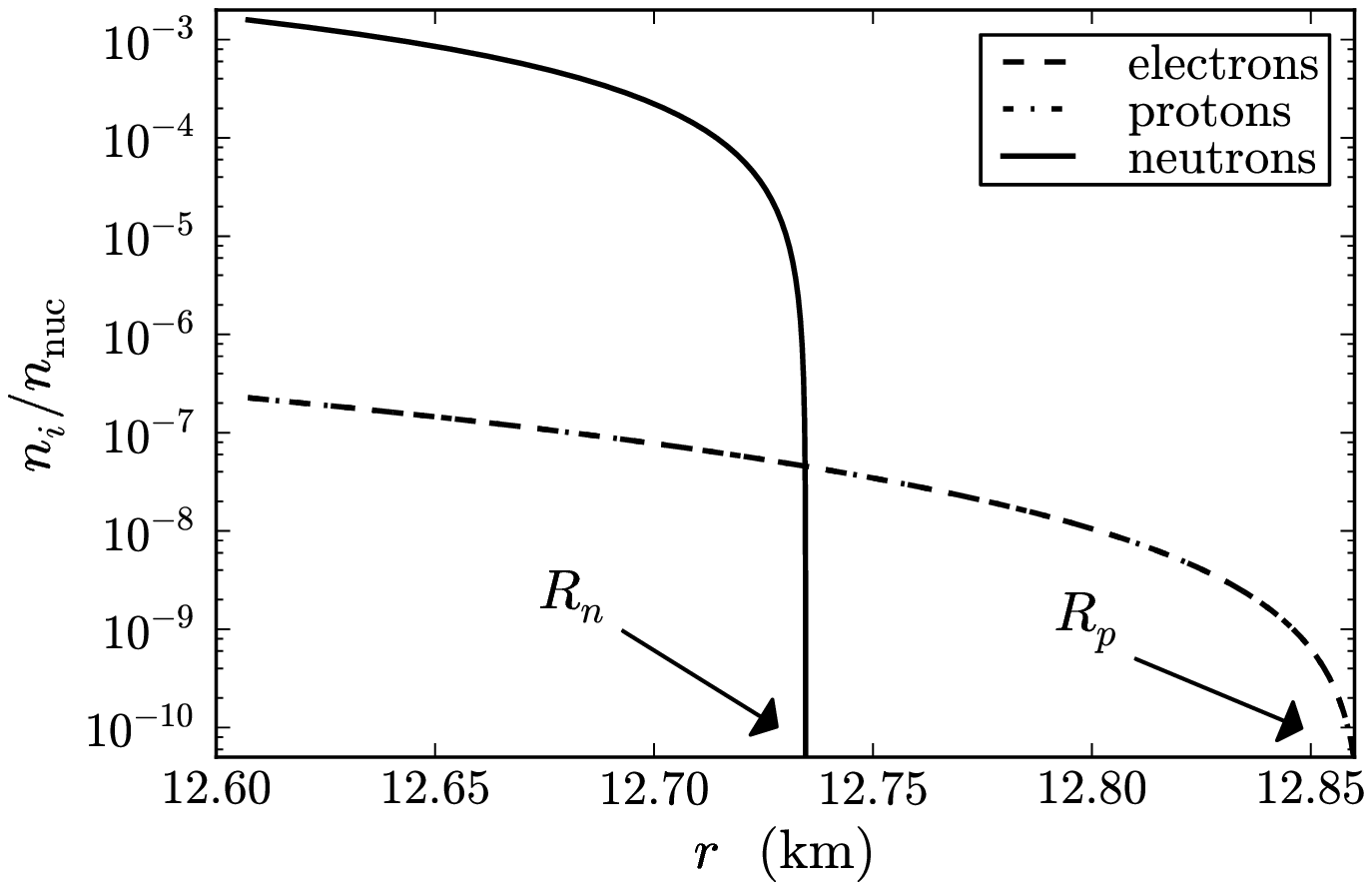} \includegraphics[width=\columnwidth,clip]{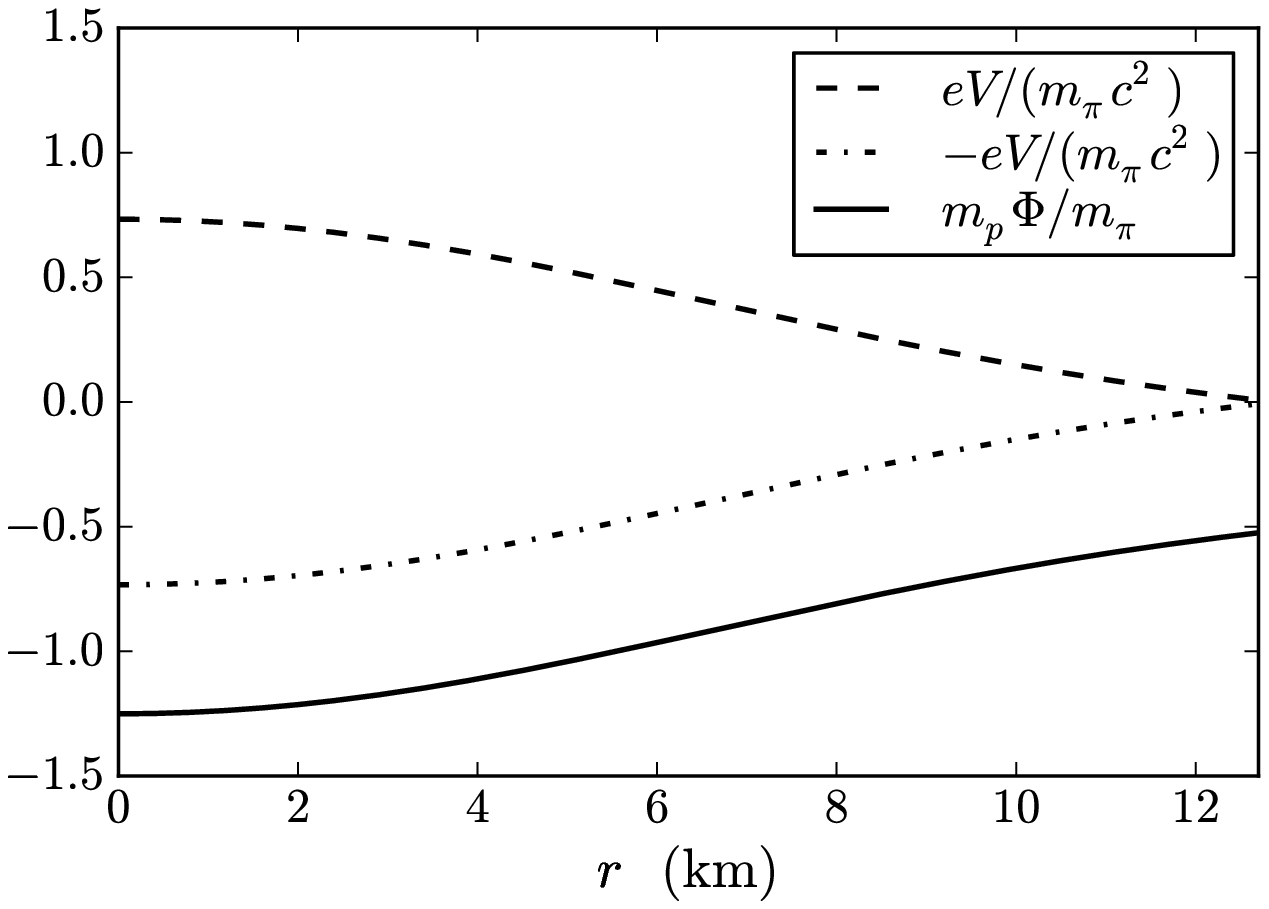}
\caption{Top panel: particle number density of neutrons, protons, and electrons approaching the boundary of the configuration in units of the nuclear density $n_{\rm nuc} \simeq 1.6\times 10^{38}$ cm$^{-3}$. Bottom panel: proton and electron Coulomb potentials in units of the pion rest-energy $eV/(m_\pi c^2)$ and $-eV/(m_\pi c^2)$ respectively and the proton gravitational potential in units of the pion mass $m_p \Phi/m_\pi$ where $\Phi=({\rm e}^{\nu/2}-1)$.}\label{fig:fig2}
\end{figure}

The solution for the particle densities, the gravitational potential, the Coulomb potential and the electric field are shown in Fig.~(\ref{fig:fig2}) for a configuration with central density $\rho(0)=3.94 \rho_{\rm nuc}$. In order to compare our results with those obtained in the case of nuclear matter cores of stellar dimensions \cite{PRC2011} as well as to analyze the gravito-electrodynamical stability of the configuration we have plotted the electric potential in units of the pion rest-energy and the gravitational potential in units of the pion-to-proton mass ratio. One particular interesting new feature is the approach to the boundary of the configuration: three different radii are present corresponding to distinct radii at which the individual particle Fermi pressure vanishes. The radius $R_e$ for the electron component corresponding to $P^F_e (R_e) = 0$, the radius $R_p$ for the proton component corresponding to $P^F_p (R_p) = 0$ and the radius $R_n$ for the neutron component corresponding to $P^F_n (R_n) = 0$. 

The smallest radius $R_n$ is due to the threshold energy for $\beta$-decay which occurs at a density $\sim 10^7$ g cm$^{-3}$. The radius $R_p$ is larger than $R_n$ because the proton mass is slightly smaller than the neutron mass. Instead, $R_e > R_p$ due to a combined effect of the difference between the proton and electron masses and the implementation of the global charge neutrality condition through the Thomas-Fermi equilibrium conditions. 

For the configuration of Fig.~\ref{fig:fig2} we found $R_n \simeq 12.735$ km, $R_p \simeq 12.863$ km and $R_e \simeq R_p + 10^3 \lambda_e$ where $\lambda_e=\hbar/(m_e c)$ denotes the electron Compton wavelength. We find that the electron component follows closely the proton component up to the radius $R_p$ and neutralizes the configuration at $R_e$ without having a net charge, contrary to the results e.g in \cite{olson78}. 

\begin{figure}
\centering
\includegraphics[width=\columnwidth,clip]{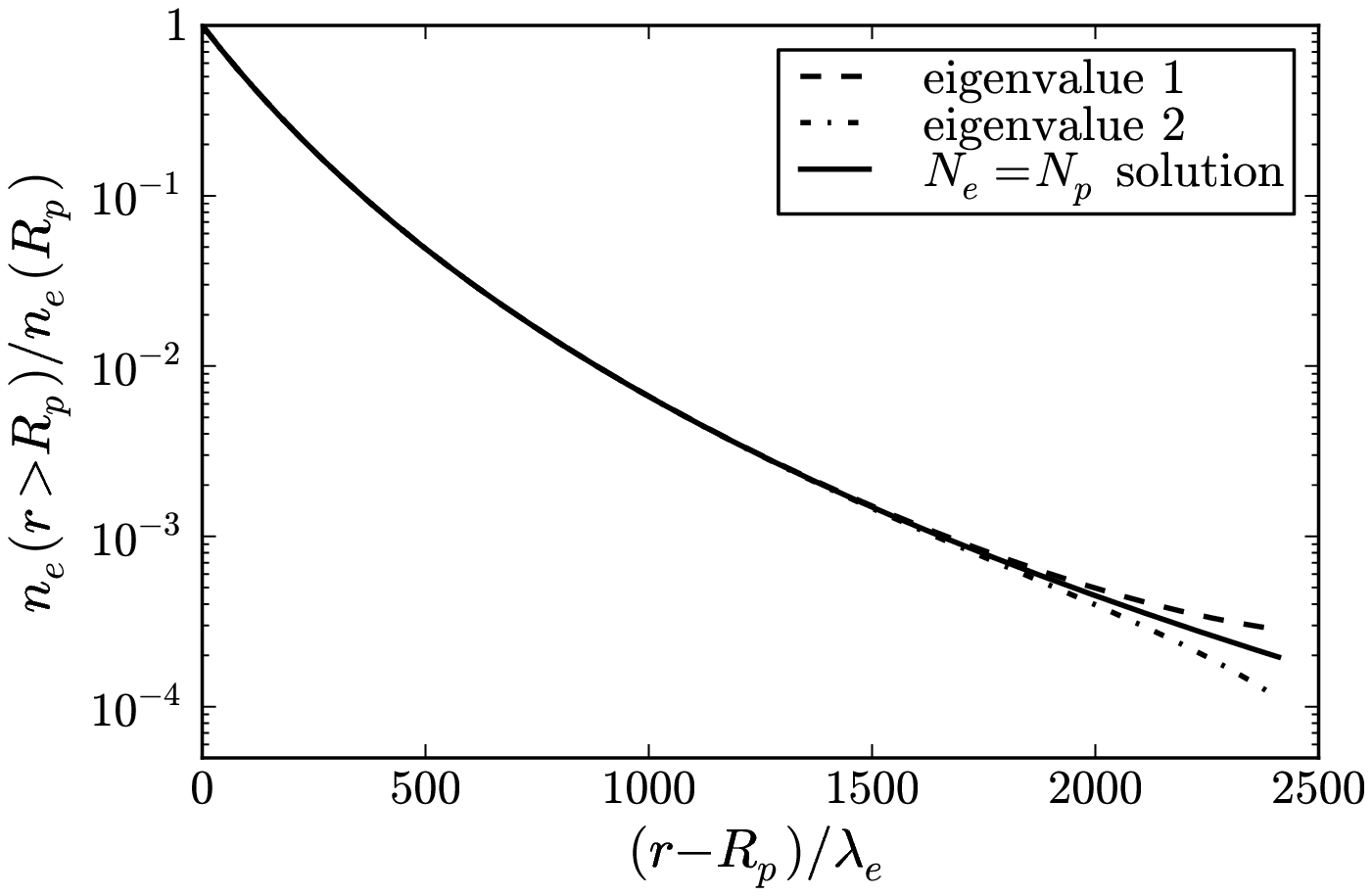} \includegraphics[width=\columnwidth,clip]{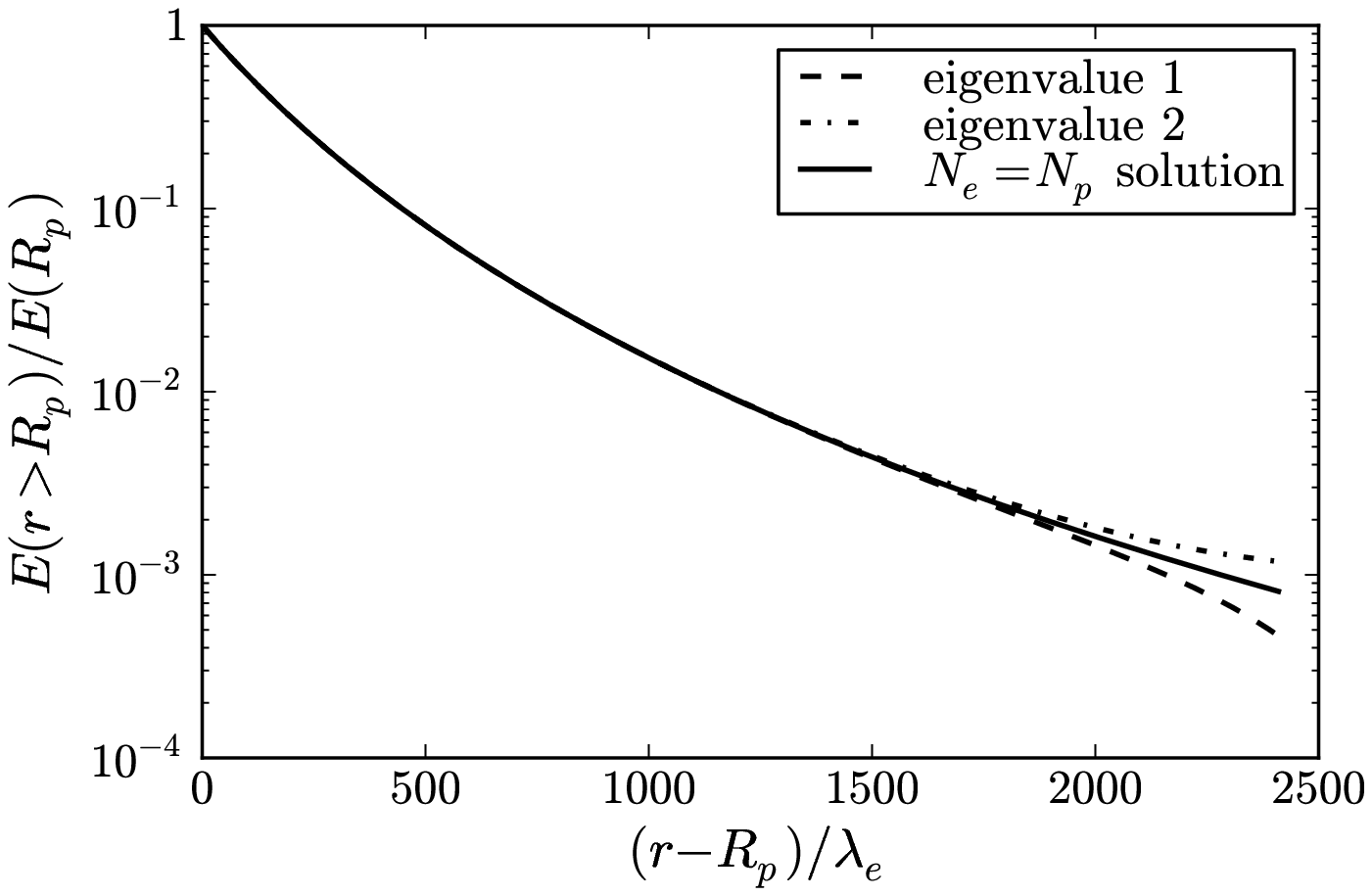}
\caption{Top panel: electron number density for $r\geq R_p$ normalized to its value at $r=R_p$. Bottom panel: electric field for $r\geq R_p$ normalized to its value at $r=R_p$. We have shown also the behavior of the solution of the general relativistic Thomas-Fermi equation (\ref{eq:GRTF}) for two different eigenvalues close to the one which gives the globally neutral configuration.}\label{fig:fig3}
\end{figure}

It can be seen from Fig.~\ref{fig:fig2} that the negative proton gravitational potential energy is indeed always larger than the positive proton electric potential energy. Therefore the configuration is stable against Coulomb repulsion. This confirms the results in the simplified case analyzed by M.~Rotondo et al. in \cite{PRC2011}.

From Eq.~(\ref{eq:olsoneq}) and the relation between Fermi momentum and the particle density $P^F_i = (3 \pi^2 \hbar^3 n_i)^{1/3}$, we obtain the proton-to-electron and proton-to-neutron ratio for any value of the radial coordinate 
\begin{eqnarray}\label{eq:nenpratio}
\frac{n_p (r)}{n_e(r)} &=& \left[ \frac{f^2(r) \mu^2_e(r)-m^2_p c^4}{\mu^2_e(r)-m^2_e c^4} \right]^{3/2}\, ,\\ 
\frac{n_p (r)}{n_n(r)} &=& \left[ \frac{g^2(r) \mu^2_n(r)-m^2_p c^4}{\mu^2_n(r)-m^2_n c^4} \right]^{3/2}\, ,
\end{eqnarray}
where $f(r)=(E^F_p + m_p c^2 - eV)/(E^F_e + m_e c^2 + eV)$, $g(r)=(E^F_p + m_p c^2 - eV)/(E^F_n + m_n c^2)$ and the constant values of the generalized Fermi energies are given by
\begin{eqnarray}\label{eq:EFi}
E^F_n &=& m_n c^2 {\rm e}^{\nu(R_n)/2}-m_n c^2\, ,\\ 
E^F_p &=& m_p c^2 {\rm e}^{\nu(R_p)/2}-m_p c^2 + e V(R_p)\, ,\\
E^F_e &=& m_e c^2 {\rm e}^{\nu(R_e)/2}-m_e c^2\, .
\end{eqnarray}

A novel situation occurs: the determination of the quantities given in Eqs.~(\ref{eq:nenpratio}) and (\ref{eq:EFi}) necessarily require the prior knowledge of the global electrodynamical and gravitational potential from the center of the configuration all the way out to the boundary defined by the radii $R_e$, $R_p$ and $R_n$. This necessity is an outcome of the solution for the eigenfunction of the general relativistic Thomas-Fermi equation (\ref{eq:GRTF}).

From the regularity condition at the center of the star $n_e(0)=n_p(0)$ together with Eq.~(\ref{eq:nenpratio}) we obtain the Coulomb potential at the center of the configuration
\begin{eqnarray}\label{eq:diffgenrel}
e V(0) &=& \frac{(m_p-m_e)c^2}{2} \Bigg[ 1 + \frac{E^F_p-E^F_e}{(m_p-m_e)c^2} \nonumber \\
&-& \frac{(m_p+m_e) c^2}{E^F_n+m_n c^2} {\rm e}^{\nu(0)} \Bigg]\, ,
\end{eqnarray}
which after some algebraic manipulation and defining the central density in units of the nuclear density $\eta = \rho(0)/\rho_{\rm nuc}$ can be estimated as
\begin{eqnarray}\label{eq:V0app}
eV(0) &\simeq& \frac{1}{2} \Bigg[ m_p c^2 {\rm e}^{\nu(R_p)/2}-m_e c^2{\rm e}^{\nu(R_e)/2} \nonumber \\
&-& \frac{m_n c^2 {\rm e}^{\nu(R_n)/2}}{1+[P^F_n(0)/(m_n c)]^2} \Bigg] \nonumber \\
&\simeq& \frac{1}{2}\Bigg[\frac{(3 \pi^2 \eta/2)^{2/3} m_p}{(3 \pi^2 \eta/2)^{2/3} m_\pi+ m^2_n/m_\pi}\Bigg] m_\pi c^2\, ,
\end{eqnarray}
where we have approximated the gravitational potential at the boundary as ${\rm e}^{\nu(R_e)/2} \simeq {\rm e}^{\nu(R_p)/2} \simeq {\rm e}^{\nu(R_n)/2} \simeq 1$. Then for configurations with central densities larger than the nuclear density we necessarily have $eV(0) \gtrsim 0.35 m_\pi c^2$. In particular, for the configuration we have exemplified with $\eta = 3.94$ in Fig.~\ref{fig:fig2}, from the above expression (\ref{eq:V0app}) we obtain $eV(0) \simeq 0.85 m_\pi c^2$. This value of the central potential agrees with the one obtained in the simplified case of nuclear matter cores with constant proton density \cite{PRC2011}.

\section{Conclusions}

We have proved in the first part of this letter that the treatment generally used for the description of neutron stars adopting the condition of local charge neutrality, is not consistent with the Einstein-Maxwell equations and microphysical conditions of equilibrium consistent with quantum statistics (see Fig.~\ref{fig:1}). We have shown how to construct a self-consistent solution for a general relativistic system of degenerate neutrons, protons and electrons in $\beta$-equilibrium fulfilling global but not local charge neutrality.

Although the mass-radius relation in the simple example considered here in our new treatment, differs slightly from the one of the traditional approaches, the differences in the electrodynamic structure are clearly very large. As is well-known these effects can lead to important astrophysical consequences on the physics of the gravitational collapse of a neutron star to a black hole \cite{physrep}.

Having established in the simplest possible example the new set of Einstein-Maxwell and general relativistic Thomas-Fermi equations, we now proceed to extend this approach when strong interactions are present \cite{pugliese2011}. The contribution of the strong fields to the energy-momentum tensor, to the four-vector current and consequently to the Einstein-Maxwell equations have to be taken into account. Clearly in this more general case, the conditions introduced in this letter have to be still fulfilled: the $r$-independence of the generalized Fermi energy of electrons and the fulfillment of the general relativistic Thomas-Fermi equation \cite{pugliese2011}. In addition, the generalized Fermi energy of protons and neutrons will depend on the nuclear interaction fields. The fluid of neutrons, protons and electrons in this more general case does not extend all the way to the neutron star surface but is confined to the neutron star core endowed with overcritical electric fields, in precise analogy with the case of the compressed nuclear matter core of stellar dimension described in \cite{PRC2011}. 




\begin{thebibliography}{}

\bibitem{physrep} R.~Ruffini, G.~V.~Vereshchagin and S.-S.~Xue, Phys.\ Rep. {\bf 487}, 1 (2010).

\bibitem{PRC2011} M.~Rotondo, Jorge A.~Rueda, R.~Ruffini and S.-S.~Xue, Phys.\ Rev.\ C {\bf 83}, 045805 (2011).

\bibitem{popov1} V.~S.~Popov, Sov.\ Phys.\ JETP\ {\bf 32}, 526 (1971).

\bibitem{popov2} Ya.~B.~Zeldovich and V.~S.~Popov, Sov.\ Phys.\ USP\ {\bf 14}, 673 (1972).

\bibitem{migdal76} 
A.~B.~Migdal, V.~S.~Popov and D.~N.~Voskresenskii, Sov.\ Phys.\ JETP Lett.\ {\bf 24} 165 (1976). 

\bibitem{migdal77} 
A.~B.~Migdal, V.~S.~Popov and D.~N.~Voskresenskii, Sov.\ Phys.\ JETP\ {\bf 45}, 436 (1977). 

\bibitem{ferreirinho80} J.~Ferreirinho, R.~Ruffini and L.~Stella, Phys.\ Lett.\ B {\bf 91}, 314 (1980).

\bibitem{ruffini81} R.~Ruffini and L.~Stella, Phys.\ Lett.\ B {\bf 102}, 442 (1981). 

\bibitem{dresden} R.~Ruffini, Proceedings of the $9^{th}$ International Conference `Path Integrals,' 
World Scientific, 207 (2008).

\bibitem{klein} O.~Klein, Rev.\ Mod.\ Phys. {\bf 21}, 531 (1949).

\bibitem{tfsol1}  G.~Scorza-Dragoni, Rend.\ Accad.\ Lincei  {\bf 8}, 301 (1928).

\bibitem{tfsol2}  G.~Scorza-Dragoni, Rend.\ Accad.\ Lincei  {\bf 9}, 378 (1929).

\bibitem{tfsol3} A.~Sommerfeld, Z.\ Physik {\bf 78}, 283 (1932).

\bibitem{tfsol4} C.~Miranda, Mem.\ Accad.\ Italia {\bf 5}, 285 (1934).

\bibitem{tfsol5} E.~Lieb and B.~Simon, Phys.\ Rev.\ Lett. {\bf 31}, 681 (1973).

\bibitem{tfsol6} E.~Lieb, Rev.\ Mod.\ Physics {\bf 53}, 603 (1981).

\bibitem{tfsol7} L.~Spruch, Rev.\ Mod.\ Physics {\bf 63}, 151 (1991).

\bibitem{olson75} E.~Olson and M.~Bailyn, Phys.\ Rev.\ D {\bf 12}, 3030 (1975).

\bibitem{olson78} E.~Olson and M.~Bailyn, Phys.\ Rev.\ D {\bf 18}, 2175 (1978).

\bibitem{kodama72} T.~Kodama and M.~Yamada, Prog.\ Theor.\ Phys. {\bf 47}, 444 (1972).

\bibitem{pugliese2011} Jorge A.~Rueda, R.~Ruffini and S.-S.~Xue, submitted to Phys.\ Lett.\ B.

\end{thebibliography}
\end{document}